# Effect of the degree of oxidation of graphene oxide on As(III) adsorption

A. C. Reynosa-Martínez[1]. G. Navarro Tovar[1], W. R. Gallegos[1], H. Rodríguez-Meléndez[1], R. Torres-García[2], G. Mondragón-Solórzano[2], J. Barroso-Flores[2], M.A. Alvarez-Lemus[3], V. García Montalvo[4], E. López-Honorato[1,*]

[1] Centro de Investigación y de Estudios Avanzados del IPN, Unidad Saltillo. Av. Industria Metalúrgica 1062, Parque Industrial. Ramos Arizpe, Coahuila, 25900, México.

[2] Centro Conjunto de Investigación en Química Sustentable UAEM-UNAM, Carretera Toluca-Atlacomulco Km 14.5, Unidad San Cayetano, Toluca, Estado de México, 50200 México.

[3] Universidad Juárez Autónoma de Tabasco, Av. Universidad s/n, Magisterial. Villahermosa, Tabasco, 86040, México.

[4] Instituto de Química. Universidad Nacional Autónoma de México, Circuito Exterior, Ciudad Universitaria, CD. MX., 04510, México.

**Abstract**

The study of the interaction between graphene oxide and arsenic is of great relevance towards the development of adsorbent materials and as a way to understand how these two materials interact in the environment. In this work we show that As(III) adsorption, primarily $H_3AsO_3$, by graphene oxide is dependent on its degree of oxidation. Variations in the concentration of potassium permanganate resulted in an increase on the C/O ratio from 1.98 to 1.35 with C-OH and C-O-C concentrations of 18 and 32%, respectively. Three oxidation degrees were studied, the less oxidized material reached a maximum As(III) adsorption capacity of 124 mg/g, whereas the graphene with the highest degree of oxidation reached a value of 288 mg/g at pH 7, to the authors knowledge, the highest reported in the literature. The interaction between graphene oxide and As(III) was also studied by Density Functional Theory (DFT) computer models showing that graphene oxide interacts with As(III) primarily through hydrogen bonds, having interaction energies with the hydroxyl and epoxide groups of 378 and 361 kcal/mol, respectively. Finally, cytotoxicity tests showed that the graphene oxide had a cellular viability of 57% with 50 μg/ml, regardless of its degree of oxidation.

*Corresponding author. Tel: 844 438-9637. E-mail: eddie.lopez@cinvestav.edu.mx (Eddie López)

## 1. Introduction

Arsenic water pollution is a global challenge that affects countries in Europe, Asia and America [1]. For example, in Latin America 14 out of 20 countries suffer of this problem [2]. In Mexico, half of the country has shown water wells contaminated with arsenic with concentrations as high as 0.624 mg/L, particularly in arid and semiarid areas where other water sources are limited [3]. These concentrations are considerably high taking into account the limit of 0.01 mg/L established by the World Health Organization.

Currently, there are different techniques designed to removed arsenic, including adsorption, which is widely used due to its simplicity of operation and scalability [4]. A wide range of adsorbent material have been used such as iron oxides, activated alumina [5], activated carbon [6] and reduced graphene oxide (rGO) with nanoparticles such as magnetite, ceria, zirconium and zirconium oxide and $Fe_3O_4/MnO_2$ [7–13]. These materials have shown maximum adsorption capacities of 14.04 up to 212.33 mg/g for As(III) species ($H_3AsO_3$, $H_2AsO_3^-$, $HAsO_3^{2-}$ and $AsO_3^{3-}$), which are the most dangerous species of As. Among these materials, graphene oxide has been used primarily to work as supporting material after reduction. Nevertheless, since graphene oxide contain oxygenated functional groups such as hydroxyl and epoxide, it has shown the capability to remove other contaminants such as uranium, plutonium and toxic dyes [14–16]. Since these functional groups are responsible for the adsorption capacity of graphene oxide, its degree of oxidation has a strong effect on the adsorption capacity of contaminants, therefore graphene oxide could be developed to adsorb high concentrations of arsenic without the use of other nanoparticles [17, 18]. This could simply the production of adsorbent materials and reduce its cost towards a practical use. Furthermore, since GO is widely used in the industry it is expected to be realized into the environment, therefore, studies on the interaction between GO and contaminants such as As, have important environmental implications as they also could help to elucidate contaminant transport processes as they have already suggested[19].

Different methods have been used to synthetize GO from graphite, one of the most known is the improved Hummers method in which $KMnO_4$ is used as oxidant agent in a mixture of $H_2SO_4:H_3PO_4$ in a ratio 9:1[19, 20]. The oxidation process starts on carbon atoms at the edges and on defects of the graphene layer, where hydroxyl groups are formed. As oxidation progress the basal plane undergoes oxidation and more hydroxyl groups are formed. Later on, the hydroxyl group on the edges are further oxidized to ketone and quinone groups, whereas the hydroxyl groups on basal plane are condensed to epoxide groups by dehydration. Finally, the ketone group is oxidized to carboxylic acid through the breakage of the carbon ring [22]. The complete oxidation of hydroxyl group to epoxide and carboxylic acid is achieved, for example, by increasing the amount of $KMnO_4$ [22, 23].

In this work we produced three different graphene oxides with a different degree of oxidation. We characterized these GOn and tested their As(III) capacity based on pH and As(III) concentration, showing maximum adsorption capacities up to 288 mg/g, to our knowledge the highest reported in the literature for single graphene oxide and even reduced graphene oxide with nanoparticles of ferrite, 147 mg/g [25]. This work also shows the effect of secondary salts on its adsorption capacity and provided computer models to explain the adsorption process and cytotoxicity to understand their impact as suitable adsorbent material. Our results show that graphene oxide can achieve high As(III) adsorption capacities, thus simplifying the possible use of graphene oxide as adsorbent material for As(III).

## 2. Experimental
### 2.1. Synthesis of graphene oxide (GO1, GO2 and GO3)

The synthesis of GO was performed using the improved Hummers method proposed by Marcano [21], as follows: 400 ml of sulfuric acid (95-98%, Jalmek) and phosphoric acid (85.8%, J.T. Beaker) were mixed in a 9:1 volume ratio with 3 g of graphite flakes (Sigma-Aldrich 95%) and 9 g of potassium permanganate (99%, Sigma-Aldrich), 1 to 3 ratio, with a weight ratio of 1:3 at 50 °C for 24 hours. After that the mixture was cooled to 2 °C and 3 ml of hydrogen peroxide (30%, Jalmek) was added. The mixture was then diluted with deionized water until pH 1. The material was washed twice with 3 different solutions, the first one with hydrochloric acid (36.5-38%, Jalmek) at 30% v/v, then with deionized water and finally with ethanol (99.5%, Jalmek). The solid material was coagulated with ethyl ether (99%, Jalmek) and centrifuged at 3500 rmp during 30 minutes (in a centrifuge XC-2450 PREMIERE). Subsequently, the material was dispersed in ethanol (99.5%, Jalmek) and exfoliated in an ultrasonic bath (Branson 3800, Frequency 40 kHz/Sonics Power 110) during 1 hour. Finally, the solid was dried overnight at 80 °C. The final product was denominated GO2.

Two graphene oxide with a lower and a higher degree of oxidation were obtained following the same procedure as described above, but with a weight ratio of 1:1 and 1:6 between graphite and potassium permanganate. This materials were designated as GO1 and GO3, respectively.

## 2.2. Characterization of graphene oxide

The solid was characterized by X-Ray Diffraction (XRD) on a PHILIPS X'Pert diffractometer using the CuKα radiation (1.5418 Å) in the 8-80° range (0.02° step size). The functional groups and carbon content were characterized by Fourier Transform-Infrared Spectroscopy (FT-IR) on a PerkinElmer Frontier FT-IR/NIR and by X-Ray Photoelectron Spectroscopy on a PHI VersaProbe II with $2 \times 10^{-8}$ mTorr vacuum chamber, aluminum anode as X-ray monochromatic source with a radiation energy of 1486.6 eV. The analysis range was from 1400 to 0 eV. The microstructure and elemental analysis were also characterized by Scanning Electron Microscopy (SEM) using a Jeol JSM-7800F Prime, Field Emission Scanning Electron Microscope. Raman Spectroscopy was also performed in a RENISHAW *inVia* Microscope using a laser excitation wavelength of 514 nm. Zeta Potential was measured in a Malvern Zetasizer Nano Z ZEN2600 using a disposable cell.

## 2.3. Adsorption tests at different pH values

To carry out the adsorption test, 0.0125 g of GOn powder were placed in a flask with 10 ml of As(III) with a concentration of 25 mg/L, the highest concentration of As recorded in nature[2, 25]. To prepare this solution an arsenic standard for ICP solution (Sigma-Aldrich, 1000 ±2 mg/L As(III) in 2% HCl, prepared with high purity $As_2O_3$) was used. The pH of the GO/As mixture was adjusted as necessary using NaOH (97%, Vetec) or HCl (36.5-38%, Sigma-Aldrich). The temperature was kept at 25 °C for 48 h, and all experiments were covered from the sun-light. At the end of the experiment all the solutions were centrifuged for 15 minutes at 3500 rpm and then filtered with a 0.45 and 0.2 μm poliethersulfone (PES) membrane. To preserve the solution 2 drops of $HNO_3$ (66.3%, J.T. Beaker) were added, according to the ASTM D2972-15 Norm [27]. The remaining As(III) in solution was quantified by Inductively Coupled Plasma Atomic Emission Spectroscopy (ICP-AES) using a PERKIN ELMER model OPTIMA 8300.

*2.4. Adsorption isotherm*

To determine the maximum adsorption capacity of GO for each oxidation degree, the previous methodology was used but with ten different As(III) concentration: 1, 25, 50, 100, 200, 300, 400, 500, 600 and 700 mg/L, in order to achieve saturation of the adsorbent material. The pH was adjusted to 7 for all the solutions, considering that the pH in natural water is in the range of 6 to 8.5 [28].

*2.5. Kinetic study*

The kinetic study was performed following the methodology described in section 2.3, but by analyzing different suspensions at 0.5, 1, 3, 6, 12, and 24 hours (see supporting information).

*2.6. Effect of natural ions in the adsorption of Arsenic(III)*

To observe the effect of natural ions on the adsorption of GOn, three ions were tested; phosphates ($PO_4^{3-}$), sulfates ($SO_4^{2-}$) and carbonates ($CO_3^{2-}$). These ions were chosen due to their abundance in natural water in arid and semi-arid regions [21–23] The concentration range for each ion was selected based on the maximum and minimum concentration reported by Navarro-Noya [32] In the case of $PO_4^{3-}$ the concentrations used were 1.5 and 30 mg/L, whereas $SO_4^{2-}$ were 170 and 1695 mg/L and $CO_3^{2-}$ were 408 and 1293 mg/L [32].

These solutions were prepared by dissolving separately $Ca_3(PO_4)_2$ (96%, Sigma-Aldrich), $CaSO_4 \cdot 2H_2O$ (98%, Sigma-Aldrich) and $CaCO_3$ (99%, Sigma-Aldrich) in deionized water, using an As(III) concentration of 25 mg/L (Sigma-Aldrich As(III) ICP standard solution) and 0.0125 g of GO. For all cases the pH was adjusted to 7. During the adsorption tests the temperature was also kept at 25 °C during 48 h. At the end of the experiment the sample was characterized as described previously.

*2.7. Computational details*

All calculations were carried out using Gaussian 09 [33]. The structures for free hosts and guests were optimized at the LC-ωPBE/6-31G(*d,p*) level of theory with the PCM continuum solvation model (*water*) and their host-guest interaction energies were calculated with the NBO deletion (NBODel) formalism as available in the NBO3.1 program supplied with the aforementioned suite. In this method, all elements in the Fock matrix which involve orbitals from host and guest simultaneously are set to zero (hence deletion); the resulting matrix is diagonalized again and the rise in energy is ascribed to the interaction energy.

*2.8. Cytotoxicity tests*

Cytotoxicity of graphene oxide materials was evaluated in mononuclear cells. Cells were cultured with RPMI 1640 medium plus 10% of Fetal Bovine Serum, 10mM Penicillin Streptomycin, and 10mM of L-glutamine (Thermofisher Scientific, USA) in a 37ºC, 5% $CO_2$ humidified incubator. Different concentration of the materials (10, 30 and 50 µg/mL) were suspended in DMSO and further incubated together with 5 x$10^3$ cells/well in a 96-well plate. After 48 h, cell viability was measured though the XTT assay (sodium 2,3,-bis(2-methoxy-4-nitro-5-sulfophenyl)-5-[(phenylamino)-carbonyl]-2H tetrazolium) according to the protocol from the supplier (Cell proliferation kit II, Roche®, Sigma-Aldrich), absorption of the dye was measured at 492 and 690 nm in an Epoch (Biotek) spectrophotometer. The data are presented as mean standard deviation of three measurements, differences between control and treated groups are shown as a result of a two-way ANOVA statistics

## 3. Results and Discussion

*3.1. Production of graphene oxide with different degrees of oxidation*

Figure 1 shows the X-ray diffraction patterns of graphite and the three types of GO produced. Graphite shows two reflections at 55 and 26°, corresponding to the (004) and (002) planes, respectively. However, after oxidation and exfoliation, GO (Figure 1 c and d) showed a single broad reflection at approximately 10° (Fig. 1a and b), which is indicative of the production of graphene oxide [21]. This change is attributed to the increase on disorder introduced by the breakage of C=C bonds and the binding of oxygenated functional groups, such as hydroxyl (–OH), carboxylic acid (COOH) and epoxide (C–O–C), which introduced defects in the carbon lattice. On the other hand, the X-ray diffraction pattern of GO1 (Figure 1b) shows two reflections at 55, 26° similar to graphite, but broaden. This could be attributed to a transition from an ordered to a disordered structure, since the reflection at 10°, similar to GO2 and GO3, is associated with the presence of a disordered structure. The presence of the three reflections is similar to other GO reported in the literature also with a low degree of oxidation [34].

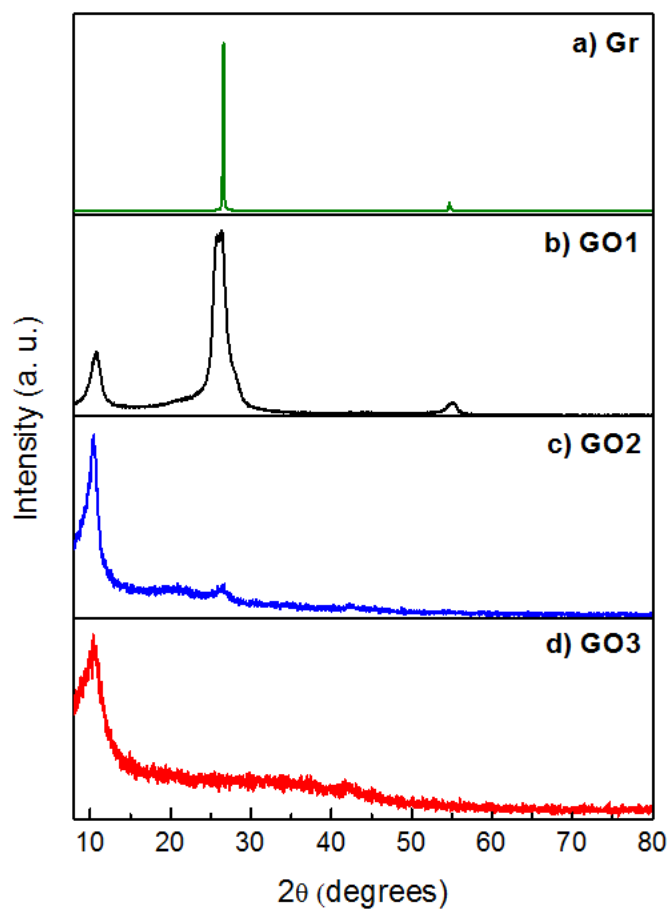

**Figure 1.** X-ray powder diffraction pattern of a) graphite, b) GO1 synthetized with a ratio of graphite and oxidant 1:1, c) GO2 synthetized with a ratio of graphite and oxidant 1:3 and d) GO3 synthetized with a ratio of graphite and oxidant 1:6.

Figure 2 (b and c) shows the FT-IR spectra of GO2 and GO3. In both cases, a broad and intense band is observed where two bands are overlapped, the first one at ~3500 cm$^{-1}$, which is attributed to stretching mode of the –OH functional group and the second ~3400 cm$^{-1}$ corresponding to the hydrogen-bonded water (H-O-H) stretching. Additionally, at ~1722 cm$^{-1}$ the C=O stretching mode, corresponding to the carboxylic acid functional group was observed. Furthermore, the peak found at ~1623 cm$^{-1}$ was assigned to the C=C stretching mode that belongs to the remaining aromatic structure of the carbon lattice. Between this two bands, GO1 presents a small signal at ~1620 cm$^{-1}$ from the vibration of the skeleton of the aromatic structure. At ~1377 cm$^{-1}$ a deformation mode band from C−H was detected, whereas at ~1225 and 1067 cm$^{-1}$ were located the bands corresponding to C−O−C and C-OH functional groups, respectively. These results confirm the oxidation of graphite and support the correct formation of GO [35]. Furthermore, Figure 2a shows the FT-IR spectra of GO1, where only a broad band at ~3500 cm$^{-1}$ attributed to stretching mode of the –OH functional group can be seen. The presence of this unique band is associated with graphite oxide and is similar to other graphene oxide reported in the literature with a low oxidation degree [18, 31].

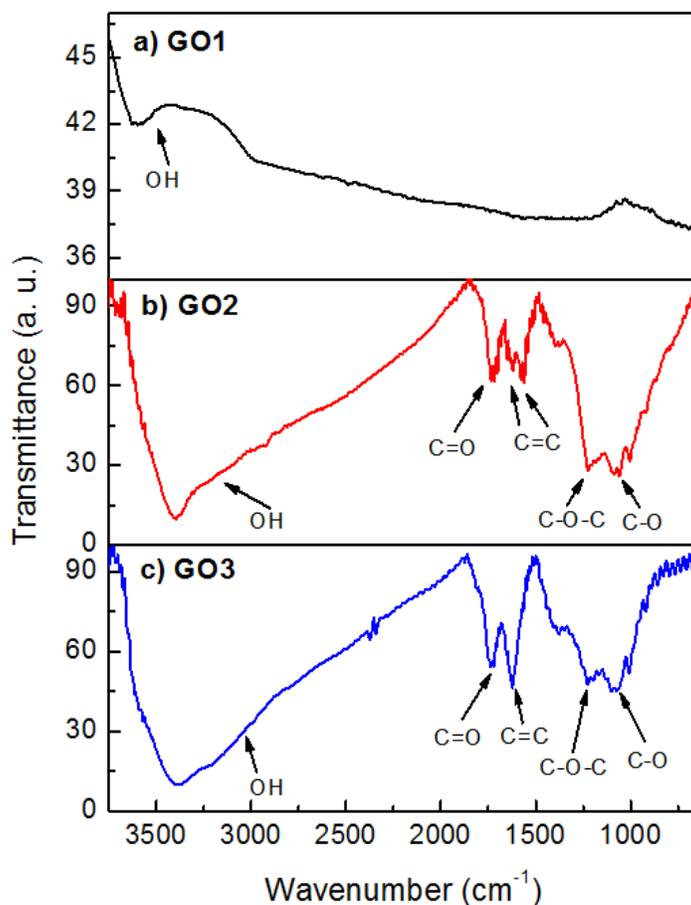

**Figure 2.** Infrared spectra of a) GO1, b) GO2 and c) GO2.

Figure 3 shows the SEM micrographs of GO1, GO2 and GO3. Figure 3 shows that the GO laminates had lengths of between 2.5 and 20 μm, with a relatively smooth surface. GO3 appeared to have similar particle sizes, but with a considerably increase on surface roughness (Fig. 3e). This difference between both materials is likely caused due to their differences on oxidation, since a higher concentration of lattice defects induce more lattice distortions that result on an increase of tortuosity of the graphene sheet, thus generating the surface roughness observed by SEM [36]. The laminar structure was corroborated also by TEM (Figure S1)

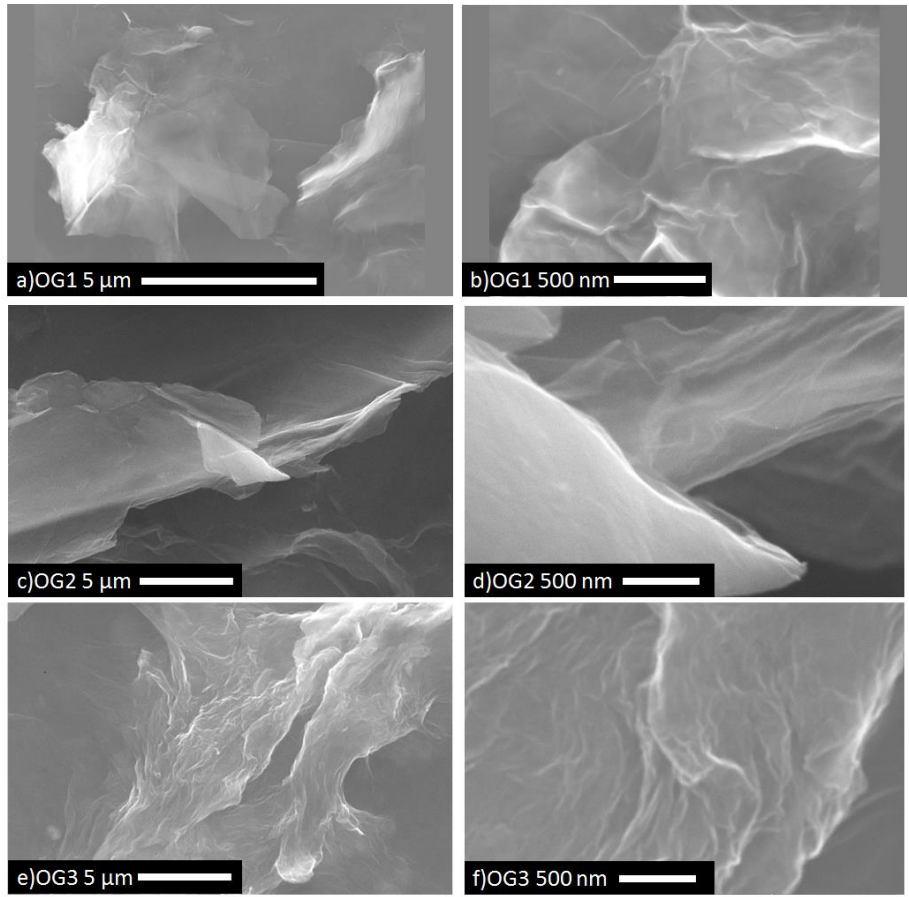

**Figure 3.** Micrograph of Scanning Electron Microscopy of a, b) GO1, c, d) GO2 and e, f) GO3.

Figure 4 shows an example of the Raman spectra of GO1 to GO3. In the Raman spectra of graphite (see supporting material) is possible to identify the G band as a narrow and intense band at ~1581 cm$^{-1}$, which originates from the bond stretching of $sp^2$ bonds, associated with the fist-order Raman mode $E_{2g}$ [37]. However, in GO this band becomes broad due to the presence of defects that affect the carbon network. Furthermore, we observed that the G band in both GO2 and GO3 shifted to 1590 cm$^{-1}$. Additionally, it was possible to identify the D band at 1356 cm$^{-1}$ for GO2 and at 1357 cm$^{-1}$ for GO3. This band is associated with the vibrational mode $A_{1g}$ that is attributed to the breathing mode of the network rings due to the defects introduced by the transformation of $sp^2$ to $sp^3$ bonds due to the attachment of oxygenated functional groups to the graphene structure [38].

Besides the G and D bands it was possible to identify the D', D'' and D* bands. These bands are attributed to the disorder introduced by the oxidation process. The D' band has been associated to the double carbon vacancy ($C_2$) in the network according to generation of consecutive rings with 5 and 8 carbon atoms [39]. On the other hand, D'' and D* bands have been associated with the content of oxygen as they shift to lower and higher wavenumbers, respectively [40]. In GO2 these bands were located at 1516 cm$^{-1}$ (D'') and 1135 cm$^{-1}$ (D*), but shifted towards 1532 cm$^{-1}$ (D'') and 1134 cm$^{-1}$ (D*) for GO3, thus suggesting a higher degree of oxidation [31, 32]. Furthermore, Figure 4a and b show the Raman spectrum of GO1, where generally two distinctive signals were obtained. Contrary to OG2 and OG3, overall, the G band (1590 cm$^{-1}$) was more intense than the D band (1359 cm$^{-1}$) for OG1, suggesting a more graphitized structure. The existence of two different Raman spectrum for OG1 suggest that the material was partially oxidized with areas maintaining a high concentration of sp2 character[42].

Furthermore, the peak intensity ratio between the D and G bands ($I_D/I_G$) is generally used as a measure of structural disorder [43]. In graphite, this ratio was 0.09, thus suggesting a material with a high degree of order, whereas GO1, GO2 and GO3 showed ratios of 0.42/0.5 (two intensity ratios as shown in Fig. 4a,b), 1.29 and 1.34 respectively, where the disorder degree increased significantly. Zhou et al. analyzed an area of 10 x10 μm in a pristine GO with a 1:3 graphite:oxidant ratio and found a homogeneous oxidation with a $I_D/I_G$ of 0.93 [44]. Vallés et al. found that at lower C/O ratio grater $I_D/I_G$ values were obtained, when C/O was ~2 the $I_D/I_G$ was 1, conversely when C/O was ~5 the $I_D/I_G$ decreased to 0.6 [45]. With this ratio it is expected that our GOn has a higher oxidation degree, and structural disorder, than reported in literature.

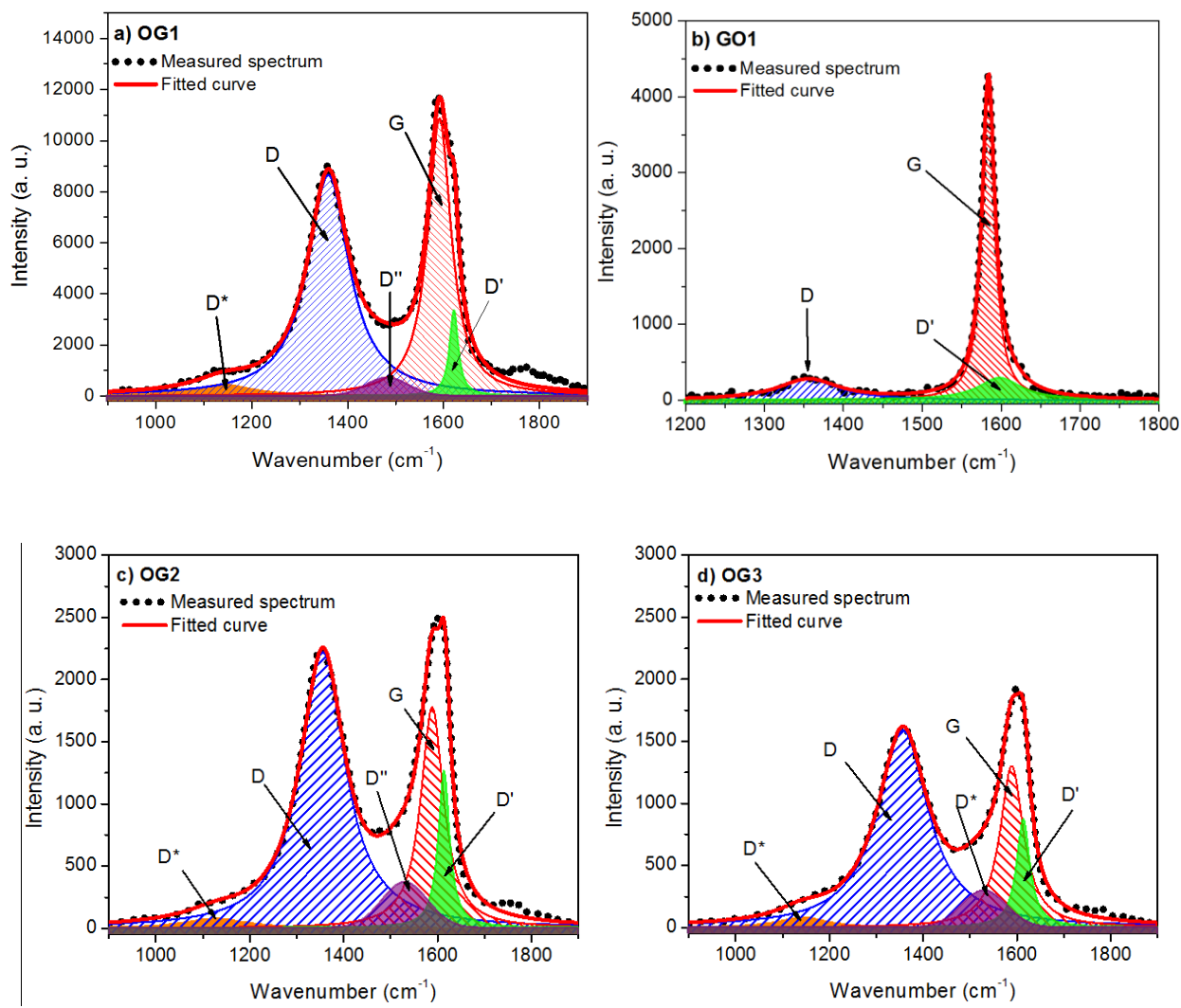

**Figure 4.** Raman spectra of primary bands of a,b) GO1, c) GO2 and d) GO3.

The three GO were also characterized by XPS as shown in Figure 5. GO1 (Figure 5a) shows an intense band at 284.7 eV attributed to graphitic structure (C=C with 54%), confirming the previous results from Raman analysis, a broad and less intensive band at 284.9 eV, attributed to *sp³* hybridization (C-C with 29.9%) generated during oxidation process, as well as the hydroxyl group at 286.9 eV (C-OH with 16.1%). The presence of only one kind of oxygenated functional group is related with the ratio of $KMnO_4$:graphite used to the synthesis of GO1 since it was not enough to oxidized the hydroxyl groups to epoxide or carboxylic acid[46]. For GO2 (Figure 5a), the C1s region (284-291eV) showed the presence of C=C (284.8 eV), C–C (285.3 eV), C–OH (286.8 eV), C-O-C (287.1 eV), C=O (288 eV) and COOH (289.6 eV) bonds, attributed to the hydroxyl, epoxide and carboxylic acid functional groups, respectively [47]. The GO2 had a C=C and C-C concentration of 19.6 and 21.7% respectively, whereas C-OH, C-O-C, C=O and COOH, had values of 5.7, 28.8, 15.4 and 8.8%, respectively. These values changed for GO3 (Figure 5b), which showed clear differences on the concentration of C=C with only 0.8%, whereas C-C, C-OH, C-O-C, C=O and COOH had a concentration of 45.7, 18, 31.7, 3 and 0.7%, respectively (See Table 1). In general, the percentage of C-OH and C-O-C in our GO3 is higher than reported for others studies where the absent functional group is –OH [25].

Overall, GO2 was composed primarily of the C-C/C=C structure (41.3%) and epoxide functional groups (28.8%). As the degree of oxidation increased, the concentration of epoxide functional groups and hydroxyl groups also increased to 31.7 and 18%, respectively. Furthermore, based on the total carbon and oxygen content, GO2 and GO3 had C/O ratios of 2.03 and 1.3, respectively. Similar percentages were reported for GO synthetized by modified Hummers method with a C/O of 2.3 and 25.33, 18.43, 41.95, 6.99 and 7.29 % of C-C/C=C, C-OH, C-O-C, C=O and COOH respectively [48], and with Hummers method was obtained a C/O ratio of 2.4 and 48.4% of C=C bond, 47.6% C-O bond and 4.7% of C=O bond. C=C bonds stem from graphitic structure, C-O was attributed to both C-O-C and C-OH functional groups while C=O was also assigned to COOH functional groups [49]. As can be seen the predominant functional groups obtained, both in literature and experimentally, were C-OH and C-O-C in contrast with C=O and COOH.

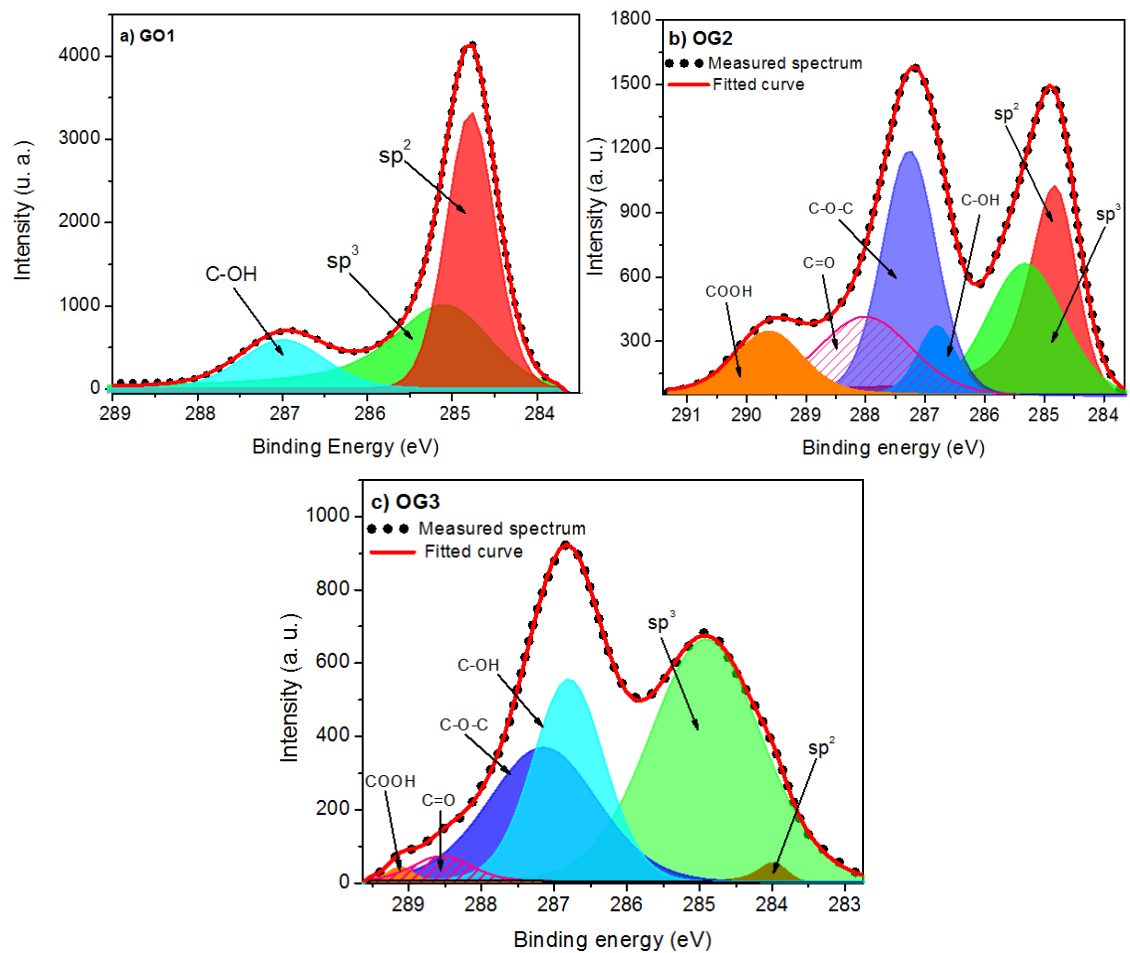

**Figure 5.** High-resolution X-ray photoelectron spectra for a) GO1, b) GO2 and c) GO3.

**Table 1.** Quantification of functional groups in GO1 and GO2 by X-ray photoelectron spectra.

| | Functional group (At %) | | | | | |
|---|---|---|---|---|---|---|
| | C=C | C–C | C–OH | C–O–C | C=O | COOH |
| OG1 | 54 | 29.9 | 16.1 | 0 | 0 | 0 |
| OG2 | 19.6 | 21.7 | 5.7 | 28.8 | 15.4 | 8.8 |
| OG3 | 0.8 | 45.7 | 18 | 31.7 | 3 | 0.7 |

Figure 6 shows the effect of pH on the values of zeta potential (ζ) of GO1 to GO3. The zeta potential of GO1 change from -36 mV at pH2 to -42 mV at pH 3 and showed de lowest value at pH 11 with

-49 mV, GO2 changed from -24 mV at pH 2, down to -35 mV at pH 3, and then decreasing with pH to -44 mV at pH 11. Similarly, GO3 had a zeta potential of -33 mV at pH 2, also decreasing with pH to -44 mV at pH 11. For most pH values, GO3 had a lower zeta potential of approximately -54 mV (81%) between pH 3 and 11 [50].

According to the literature, Zeta Potential (ζ) values above +30 mV ("more positive") and below -30 mV ("more negative") suggest a stable suspension at a specific pH value [51]. The three materials got ζ values below -30 mV (except GO2 at pH 2), thus suggesting that highly oxidized GO is stable at any pH, but most stable at alkaline pH values. The stability in suspension of the material is attributed to the concentration of oxygenated functional groups and its ionization. Krishnamoorthy, et al. associated the most negative ζ values to the dissociation of acid functional groups such as carboxylic acid ($H^+ + {}^-O–C=O$) and hydroxyl ($O^- + {}^+H$) [34]. The stability of GO1 is attributed to its partial oxidation and the predominance of −OH groups, whose negative charge is enhanced with the increase of pH. Even in low concentration, this oxygenated group, provides a stability in suspension [52]. The low concentration of acid groups in GO2 is associated with its stability in solution since GO3 has a higher density of these groups as confirmed by XPS analysis. Baskoro et al. reported a GO prepared with a graphite:oxidant ratio 1:1 with a ζ of -24 mV pH=3, -38 mV at pH=10, and between -31.5 to -33 mV at pH=7, they attributed this phenomena to the ionization to the carboxylic acid groups even though their GO had a 5% of COOH groups but 45% of O-C-O functional group (according to the XPS results) [53]. The extremely low concentration in COOH group, and the absence of C-OH, could be the reason for a lower ζ compared with our GOn. On the other hand, Shih, et al. reported a GO with a ratio 1:3 unstable at acid pH with -4.25 mV (pH=1) and high stability at pH=14 with -44.73 mV attributed to the ionization of functional groups, especially to COOH in the layer edges[54].

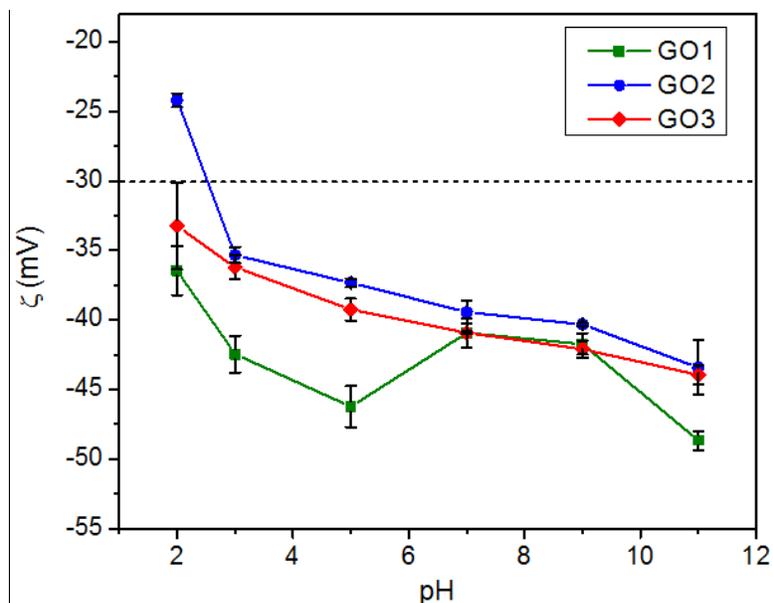
**Figure 6.** Zeta Potential plot at different pH values.

*3.2. Adsorption tests at different pH values*

Figure 7 shows the effect of pH on the adsorption capacity (*q*) of arsenic on GOn. It shows that *q* is dependent on pH, for example, GO1 have an adsorption capacity of 5.7, 4.4 and 2.4 mg/g at pH pH 3, 7 and 11, respectively. Similar behavior was observed for GO2 that reduces its adsorption capacity from 5.4 mg/g at pH 3, down to 4.5 at pH 7 and 1.4 at pH 11. As for GO3, its adsorption capacity appears to be slightly lower than GO2, particularly at pH 3, where it had a *q* of 3.7 mg/g. In the range of pH 7-9, the As adsorption of GO1, GO2 and GO3 was very similar, having identical values of *q* of 4.6 and 4 mg/g at pH 7 and 9 (Figure 7). Nevertheless, As(III) adsorption appears to increase with lower oxidation degrees particularly at pHs below 5 and above 11 (outside the range of pH found in natural water).

The higher adsorption capacity of GOn at acidic conditions is related to the oxygenated functional groups attached to the *sp²* and *sp³* carbon atoms. Unlikely alkaline environments, at acid pH values, the oxygenated functional groups have not been dissociated, thus are not negatively charged, making it possible to interact with arsenic despite having a neutral charge ($H_3AsO_3$) [55]. On the other hand, as pH values increase, particularly at pH ≥11, both the oxygenated functional groups and the arsenic species are dissociated into ⁻O–C=O, –O⁻, and $H_2AsO_3^-$, $HAsO_3^{2-}$ and $AsO_3^{3-}$ [56], therefore, are negatively charged, thus reducing the adsorption process due to repulsion. Similarly, the almost identical $q_e$ of GO1, GO2 and GO3 at pH 7 and 9 could be attributed to the incomplete dissociation of oxygenated functional groups, in the presence of the predominant specie $H_3AsO_3$.

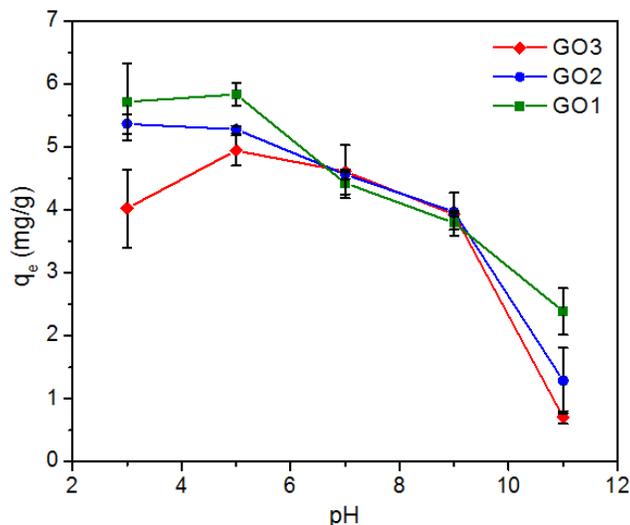

**Figure 7.** Equilibrium adsorption capacity of GO2 and GO3 at different pH values.

*3.3. Adsorption isotherm*

Figure 8 shows the relationship of adsorption at equilibrium ($q_e$) and equilibrium concentration ($C_e$) up to 550 mg/L of As (III). Figure 8 shows that GO2 and GO3 have relatively similar adsorption behavior at $C_e$ below 100 mg/L. For example, at 50 mg/L GO2 and GO3 had a $q_e$ of 8.89 and 8.90 mg/g, respectively, while GO1 seems to have a $q_e$ slightly higher of 11.6 mg/g. However, as the concentration increased, GO3 showed a higher adsorption capacity, for example at $C_e$ of 550 mg/LGO1 reached values of $q_e$ of 90 mg/g and GO2 had a $q_e$ of 178 mg/g, whereas GO3 had values of 196 mg/g. The values of the adsorption isotherms were used to study the kinetics of the adsorption process.

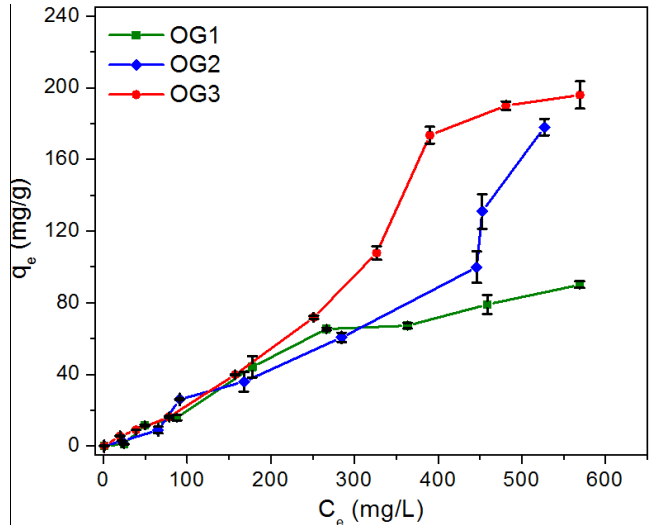

**Figure 8.** Equilibrium adsorption capacity plot against As(III) concentration at equilibrium.

The adsorption isotherm describes the interaction between adsorbent material and the adsorbate, explaining the adsorbate molecules distribution between bulk solution and adsorbent material when both phases reach equilibrium [57]. There are two adsorption isotherms often used to study adsorption, Langmuir and Freundlich. The first one assumes a homogeneous distribution of active sites and the lack of interaction between adsorbed molecules, thus proposing that the adsorbate tends to form a monolayer [58]. The equation to describe the Langmuir's isotherm model is the following[59]:

$$\frac{C_e}{q_e} = \frac{1}{Q_0 b} + \left(\frac{1}{Q_0}\right) C_e \qquad (1)$$

Where $C_e$ is the equilibrium concentration of adsorbate expressed in mg/L, $q_e$ is the adsorption capacity at equilibrium (the amount of adsorbate adsorbed per unit mass of adsorbent material) expressed in mg/g and $Q_0$ and $b$ are Langmuir constants. In order to find out if As(III) adsorption on GO1 and GO2 followed the Langmuir isotherm, the ratio $C_e/q_e$ was plotted against $C_e$ to obtain the correlation coefficient, $R^2$. The GO1, GO2 and GO3 $R^2$ was 0.7378, 0.3156 and 0.6101, respectively, thus suggesting that As(III) adsorption on graphene oxide does not follow a Langmuir behavior [60].

The second isotherm model generally used to study adsorption processes is the Freundlich model, which considers the existence of an heterogeneous surface with different energies and the formation of a multilayer of adsorbate [58]. This model follows the following logarithmic expression [59]:

$$log\ q_e = log k_F + \left(\frac{1}{n}\right) log C_e \qquad (2)$$

Where $k_F$ and $n$ are the Freundlich constants. In order to evaluate the applicability of Freundlich´s isotherm model to GO adsorption process, $log\ q_e$ against $log\ C_e$ was plotted in Figure S4 in supporting information. The correlations $R^2$ found were 0.9322, 0.9818 and 0.9917 for GO1, GO2 and GO3, respectively, thus suggesting that As(III) adsorbs on graphene oxide following the Freundlich model [61]. This result deviates from previous reports suggesting a Langmuir´s adsorption model, which assumes the same energy for active sites all over the material. However, XPS analysis have shown the existence of a diverse concentration of functional groups, specially hydroxyl and epoxide, therefore it is unlikely that all of these functional groups will interact with the same energy.

Furthermore, once obtained $k_F$ and $n$ (Table 2) the maximum adsorption capacities were calculated for each material using equation 3. The maximum adsorption capacities for GO1, GO2 and GO3 were 123, 175 and 288 mg/g, values for GO2 and GO3 are considerably higher from those previously reported for GO of 19 mg/g and even higher than some reduced graphene oxide/nanoferrites of 147 mg/g [25]. The increase of adsorption capacity from 175 to 288 mg/g suggest that the degree of oxidation and the type of functional groups on graphene oxide have a strong influence on the adsorption capacity of GO. Therefore, differences during its synthesis will have a strong effect on the adsorption of this material, suggesting that GO will not have a single generic maximum adsorption capacity that can be assigned to this material [62]. Adsorption capacity like this has not been reported before for the use of pristine GO to the removal of As(III), neither have been used Freundlich's model to describe the adsorption process, this work may help to understand the interaction between As species and GO functional groups in addition to the interaction of both with the anions from natural ground water.

$$k_F = \frac{q_m}{(C_e)^{\frac{1}{n}}} \tag{3}$$

Table 2. Freundlich constants and maximum adsorption capacity for GO1 and GO2.

| Material | $k_F$ | $n$ | $R^2$ | $q_m$ |
|---|---|---|---|---|
| GO1 | 0.5614 | 1.1760 | 0.9322 | 123 mg/g |
| GO2 | 0.1528 | 0.9475 | 0.9818 | 175 mg/g |
| GO3 | 0.0904 | 0.8125 | 0.9917 | 288 mg/g |

*3.4. Effect of natural anions in the adsorption of Arsenic (III)*

Ions such $SO_4^{2-}$, $CO_3^{2-}$ and $PO_4^{3-}$ have a strong effect on the adsorption capacity of materials to the point that they can reduce beyond 50% the adsorption capacity of nanomaterials [63]. The effect of these ions is even more relevant considering that in arid and semi-arid regions the underground waters contain high concentrations of these ions [32]. Therefore, it is necessary to study the effect of $SO_4^{2-}$, $CO_3^{2-}$ and $PO_4^{3-}$ on GO to adsorb arsenic. Figure 10 shows the effect of such ions on the adsorption capacity of GO. Only the maximum and minimum concentrations of these ions were used considering the values reported by Navarro-Noya [32]. As observed in Figure 9, $PO_4^{3-}$ (1.5-30 mg/L) has a minimum effect on $q_e$, since the adsorption reduced from 5.6 down to 4.15 mg/g with 30 mg/L of $PO_4^{3-}$. On the other hand, $SO_4^{2-}$, showed a clear effect on the adsorption of As(III) since at 1.25 mg/g it induced a reduction on adsorption capacity of 22 %, as the values changed from 5.6 mg/g down to 4 mg/g for 1695 mg/L $SO_4^{2-}$. Furthermore, $CO_3^{2-}$ showed the strongest effect on GO since at 200 and 1290 mg/L the adsorption of As(III) reduced to 2.3 and 1.7 mg/g, a reduction of 41 and 30% than the original value, respectively.

Although Wei Wan and co-workers reported that concentration between 1-4 mg/L inhibited As removal using Lepidocrocite (γ-FeOOH), they attribute this effect to $PO_4^{3-}$ competing with arsenic species for the active sites on the material surface [64]. Furthermore, even when the $SO_4^{2-}$ concentration was higher than $PO_4^{3-}$ (170-1695 mg/L), its effect on $q_e$ was minimum since the $q_e$ without any anion was 4.5 mg/g, whereas after the addition of $SO_4^{2-}$ the $q_e$ reduced to 3.05 ±0.99 and 3.61 ±0.46 mg/g for 170 and 1695 mg/L of sulfates. Interestingly, the highest concentrations of sulfates showed a higher adsorption capacity $q_e$. This effect could be attributed to the As affinity towards sulfur groups like $SO_4^{2-}$ which could be favoring adsorption process, however its effects merit further studies to clearly explain this result [63]. In the case of $CO_3^{2-}$, which had the strongest effect at all concentrations, it is likely that affected As adsorption by saturating the GO surface, thus inhibiting As-GO interaction [65].

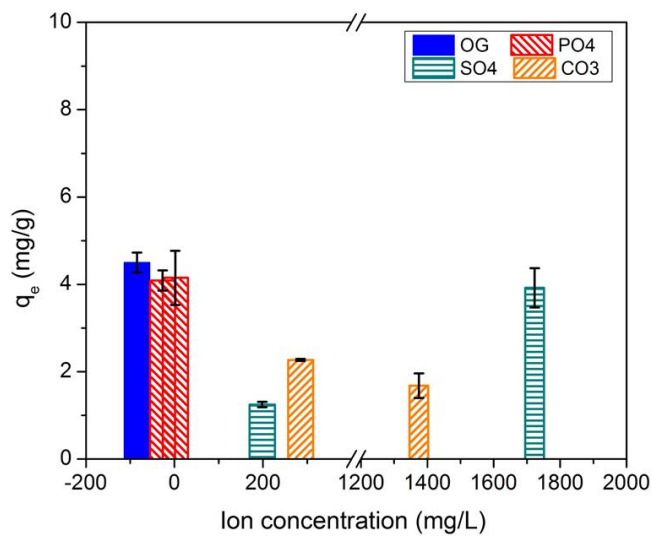

**Figure 9.** Ion effect on adsorption capacity for GO3. Arsenic initial concentration 25 mg/L.

*3.5. Computational simulation*

A striking difference in the interaction energy between the arsenic containing species and the various functional groups in GO is observed, ranging from less than 1.0 to over 380.0 kcal/mol. Table 3 gathers the calculated interaction energies between the arsenic guests and the corresponding functional group to which it is bound in GO.

**Table 3.** Interaction Energies [kcal/mol] calculated at the LC-ωPBE/6-31G(*d,p*) level of theory.

| Functional group | As (III) | | |
|---|---|---|---|
| | $H_3AsO_3$ | $H_2AsO_3^-$ | $HAsO_3^{2-}$ |
| **Graphene** | 3.90 | 254.54 | 383.35 |
| **Epoxide** | 360.57 | 329.98 | 0.72 |
| **Hydroxyl** | 378.50 | 383.35 | 378.98 |

The bare graphene sheet interacts strongly with the charged arsenic species possibly due to a strong polarization effect of the π electron density. Hydroxyl groups strongly interact with all species charged or neutral through the formation of hydrogen bonds. Finally, epoxide groups are the least favored functional group throughout the list of arsenic species in table 3. Frontier orbitals HOMO and LUMO are, in all cases, localized over the π system of graphene regardless of the arsenic species adsorbed onto it, as illustrated in Figure 10, which leads to the conclusion that their interaction is mostly electrostatic in nature. A full gathering of frontier orbitals is available in Table S6 of the supporting information section.

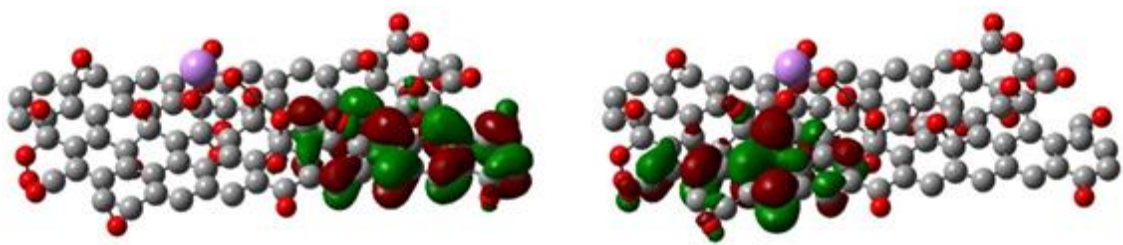

**Figure 10.** Frontier HOMO (left) LUMO (right) orbitals for the $H_3AsO_3$ adsorbed onto graphene oxide calculated at the LC-ωPBE/6-31G(*d,p*) level of theory (hydrogen atoms omitted for the sake of clarity).

To further address the nature of the graphene–arsenous acid interaction, the molecular electrostatic potential (MEP) was calculated and mapped onto the electron density isosurface; distribution of the electrostatic potential is dependent of the charge of the arsenic species adsorbed, yet it is only locally so, the potential on graphene remains mostly undisturbed except around the region on which the adsorption takes place. Figure 11 shows a comparison between the adsorption of the neutral $H_3AsO_3$ compound and the anionic $HAsO_3^{2-}$, the latter increases the negative potential (red zones) around the adsorbed arsenic compound whereas the former exhibits a more neutral potential distribution. All MEP maps are available in the supporting information in Table S7.

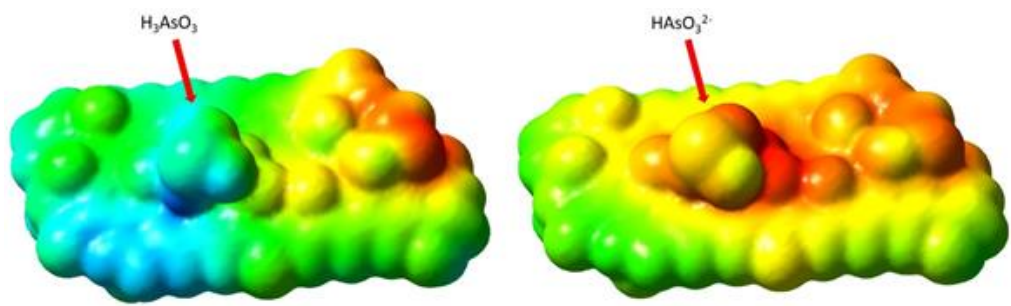

**Figure 11.** Molecular Electrostatic Potential (MEP) mapped onto the electron density isosurface calculated at the LC-ωPBE/6-31G(*d,p*) level of theory.

*3.6. Cytotoxicity of Graphene Oxide*

If graphene oxide is to be used as adsorbent material or for other applications, is important to ensure that this material will not represent an environmental hazard. Therefore, we have characterized the cytotoxicity of the three GO produced. Figure 12 shows the dose-dependent effect of GO in cellular viability, where monocytes were used as test cells. At 10 μg/ml the cellular viability decreased by 20% compared to the control group (only DMSO), while at 50 μg/ml cellular viability was reduced by 45%.

To test the toxicity of GO Farid Ahmed, et al. used diverse microbial communities from a wastewater treatment plant with a higher dosage of GO, from 10 to 300 mg/L. They found that all concentration inhibited the bacterial metabolic activity, especially those exposed to concentration between 100 to 300 mg/L, the reduction was between 50 and 70% and caused a decrease in bacterial growth of 35%[66]. Xiangang Hu et al. studied not only the phytotoxicity of GO in wheat but also the toxicity of arsenic as an aggregate. The concentration of GO was from 0.1 to 10 mg/L and As was 10 mg/L. Separately neither GO nor As caused a significant inhibition on seed germination or plan growth but together, even at 0.1 mg/L, caused a reduction of shoot length and low chlorophyll content [19].

The toxicity of GO and others carbon based materials are attributed to its high content of oxygenated functional groups and its small particle size, since this characteristic allow them to interact with cells and generate membrane damage and oxidative stress [67]. Due to the high concentration of oxygenated functional groups both GO showed a significate inhibition in cellular viability, in spite of the different oxidation degree.

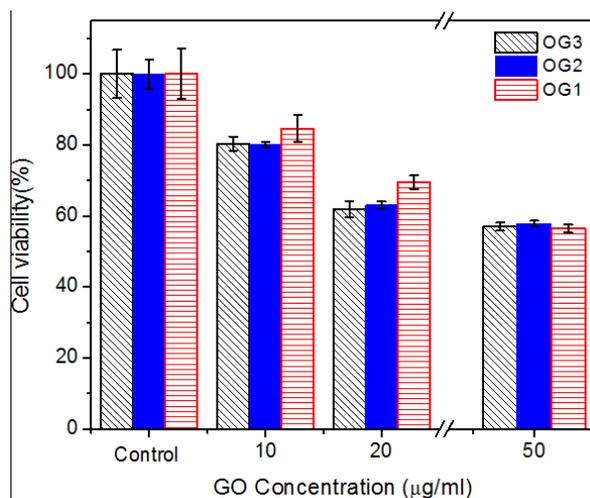

**Figure 12.** Cytotoxicity of graphene oxide with different degrees of oxidation.

## 4. Conclusions

Our work has shown that the adsorption capacity of GO can be tuned towards higher values, in this case for As(III), by controlling the degree of oxidation of the material. As the oxidation degree increased from 1.98 to 1.35 C/O ratios through the increase of $KMnO_4$ during the synthesis process, the As(III) maximum adsorption capacity went from 123 to 288 mg/g, to the authors knowledge the highest adsorption capacity reported for OG, and among the highest reported for other nanomaterials. Experimental and modelling work demonstrated that the presence of hydroxyl and epoxide groups were important to adsorb As(III) through hydrogen bonds as they could reach interaction energies of 370 and 361, respectively. Nevertheless, our results show that pH and As(III) concentration play an important role on the adsorption of As(III) since at pH 7, no clear difference between GO with different degrees of oxidation was detected. Furthermore, our work also stressed the importance of considering the cytotoxicity and the influence of secondary salts towards the development of adsorbent materials. Cell viabilities of 57% at 50 µg/L were measured for GO regardless of the degree of oxidation, whereas the presence of $CO_3^{2-}$ can reduce the adsorption capacity of As(III) almost by 30%. Therefore, further improvements need to be made towards the development of an As(III) graphene oxide based material with high adsorption capacity, low cytotoxicity and high selectivity, particularly for its used in hard waters.

**Acknowledgements**

This material is based upon work supported by a grant from the Consejo Nacional de Ciencia y Tecnología (CONACYT, project number: 247080, Problemas Nacionales). The authors would like to acknowledge CONACYT for the PhD and MSc grant awarded to A.C. Reynosa Martinez, G. Navarro Tovar and W.R. Gallegos Perez, respectively.

**References**

[1] R. Brunt, L. Vasak, and J. Griffioen, "Arsenic in groundwater: Probability of occurrence of excessive concentration on global scale," Utrecht, 2004.

[2] J. Bundschuh *et al.*, "One century of arsenic exposure in Latin America: A review of history and occurrence from 14 countries," *Sci. Total Environ.*, vol. 429, pp. 2–35, 2012.

[3] F. I. Arreguín-Cortés, R. Chávez-Guillén, and P. R. Soto-Navarro, "subterránea en México," *Tláloc, Asoc. Mex. Hidráulica*, vol. 45, no. 45, p. 11, 2009.

[4] M. Bissen and F. H. Frimmel, "Arsenic - A review. Part II: Oxidation of arsenic and its removal in water treatment," *Acta Hydrochim. Hydrobiol.*, vol. 31, no. 2, pp. 97–107, 2003.

[5] S. Goldberg, "Competitive Adsorption of Arsenate and Arsenite on Oxides and Clay Minerals," *Soil Sci. Soc. Am. J.*, vol. 66, no. 2, p. 413, 2002.

[6] D. Mohan and C. U. Pittman, "Arsenic removal from water/wastewater using adsorbents- A critical review," *J. Hazard. Mater.*, vol. 142, no. 1–2, pp. 1–53, 2007.

[7] K. Zhang, V. Dwivedi, C. Chi, and J. Wu, "Graphene oxide/ferric hydroxide composites for efficient arsenate removal from drinking water," *J. Hazard. Mater.*, vol. 182, no. 1–3, pp. 162–168, 2010.

[8] T. S. Sakthivel, S. Das, C. J. Pratt, and S. Seal, "One-pot synthesis of a ceria-graphene oxide composite for the efficient removal of arsenic species," *Nanoscale*, vol. 9, no. 10,

pp. 3367–3374, 2017.

[9] T. Y. Lin and D. H. Chen, "One-step green synthesis of arginine-capped iron oxide/reduced graphene oxide nanocomposite and its use for acid dye removal," *RSC Adv.*, vol. 4, no. 56, pp. 29357–29364, 2014.

[10] H. J. Shipley, S. Yean, A. T. Kan, and M. B. Tomson, "Adsorption of arsenic to magnetite nanoparticles: Effect of particle concentration, pH, ionic strength, and temperature," *Environ. Toxicol. Chem.*, vol. 28, no. 3, pp. 509–515, 2009.

[11] S. R. Chowdhury and E. K. Yanful, "Arsenic removal from aqueous solutions by adsorption on magnetite nanoparticles," *Water Environ. J.*, vol. 25, no. 3, pp. 429–437, 2011.

[12] S. Pourbeyram, S. Alizadeh, and S. Gholizadeh, "Simultaneous removal of arsenate and arsenite from aqueous solutions by graphene oxide-zirconium (GO-Zr) nanocomposite," *J. Environ. Chem. Eng.*, vol. 4, no. 4, pp. 4366–4373, 2016.

[13] D. Cortés-Arriagada and A. Toro-Labbé, "Improving As(III) adsorption on graphene based surfaces: Impact of chemical doping," *Phys. Chem. Chem. Phys.*, vol. 17, no. 18, pp. 12056–12064, 2015.

[14] Z. Wang, R. T. Bush, and J. Liu, "Arsenic(III) and iron(II) co-oxidation by oxygen and hydrogen peroxide: Divergent reactions in the presence of organic ligands," *Chemosphere*, vol. 93, no. 1, pp. 1936–1941, 2013.

[15] Q. Wu, J. Lan, C. Wang, Y. Zhao, Z. Chai, and W. Shi, "Understanding the Interactions of Neptunium and Plutonium Ions with Graphene Oxide: Scalar-Relativistic DFT Investigations," *J. Phys. Chem. Lett.*, vol. 118, pp. 10273–10280, 2014.

[16] G. K. Ramesha, A. Vijaya Kumara, H. B. Muralidhara, and S. Sampath, "Graphene and graphene oxide as effective adsorbents toward anionic and cationic dyes," *J. Colloid Interface Sci.*, vol. 361, no. 1, pp. 270–277, 2011.

[17] G. Z. Kyzas, E. A. Deliyanni, and K. A. Matis, "Graphene oxide and its application as an adsorbent for wastewater treatment," *J. Chem. Technol. Biotechnol.*, vol. 89, no. 2, pp. 196–205, 2014.

[18] S. Thangavel and G. Venugopal, "Understanding the adsorption property of graphene-oxide with different degrees of oxidation levels," *Powder Technol.*, vol. 257, pp. 141–148, 2014.

[19] X. Hu, J. Kang, K. Lu, R. Zhou, L. Mu, and Q. Zhou, "Graphene oxide amplifies the phytotoxicity of arsenic in wheat," *Sci. Rep.*, vol. 4, pp. 1–10, 2014.

[20] W. S. Hummers and R. E. Offeman, "Preparation of Graphitic Oxide," *J. Am. Chem. Soc.*, vol. 80, no. 6, p. 1339, 1958.

[21] D. C. Marcano *et al.*, "Improved synthesis of graphene oxide," *ACS Nano*, vol. 4, no. 8, pp. 4806–4814, 2010.

[22] Z. Liu, X. Duan, X. Zhou, G. Qian, J. Zhou, and W. Yuan, "Controlling and formation mechanism of oxygen-containing groups on graphite oxide," *Ind. Eng. Chem. Res.*, vol. 53, no. 1, pp. 253–258, 2014.

[23] T. Szabó *et al.*, "Evolution of Surface Functional Groups in a Series of Progressively Oxidized Graphite Oxides," *Chem. Mater.*, vol. 18, no. 11, pp. 2740–2749, 2006.

[24] G. Shao, Y. Lu, F. Wu, C. Yang, F. Zeng, and Q. Wu, "Graphene oxide: The mechanisms of oxidation and exfoliation," *J. Mater. Sci.*, vol. 47, no. 10, pp. 4400–4409, 2012.

[25] H. Su, Z. Ye, and N. Hmidi, "High-performance iron oxide–graphene oxide nanocomposite adsorbents for arsenic removal," *Colloids Surfaces A Physicochem. Eng. Asp.*, vol. 522, pp. 161–172, 2017.

[26] D. Van Halem, S. A. Bakker, G. L. Amy, and J. C. Van Dijk, "Arsenic in drinking water: A worldwide water quality concern for water supply companies," *Drink. Water Eng. Sci.*, vol. 2, no. 1, pp. 29–34, 2009.

[27] American Society for testing and Materials, "ASTM D2972-15 Standard Test Methodos for Arsenic in Water." pp. 1–10.

[28] F. Fernández-Luqueño, "Physicochemical and microbiological characterization for drinking water quality assessment in Southeast Coahuila, Mexico," *Int. J. Environ. Pollut.*, vol. 59, pp. 78–92, 2016.

[29] M. A. Armienta, R. Rodriguez, A. Aguayo, N. Ceniceros, G. Villaseñor, and O. Cruz, "Arsenic contamination of groundwater at Zimapán, Mexico," *Hydrogeology Journal*, vol. 5, no. 2. pp. 39–46, 1997.

[30] I. Rosas, R. Belmont, A. Armienta, and A. Baez, "Arsenic concentrations in water, soil, milk and forage in Comarca Lagunera, Mexico," pp. 133–149, 1997.

[31] G. Hernández Ordáz, M. A. Segura Castruita, L. C. Álvarez González Pico, R. A. Aldaco Nuncio, M. Fortis Hernández, and G. González Cervantes, "Comportamiento del arsénico

en suelos de la región lagunera de Coahuila, México," *Terra Latinoam.*, vol. 31, no. 4, pp. 295–303, 2013.

[32] Y. E. Navarro-Noya *et al.*, "Pyrosequencing Analysis of the Bacterial Community in Drinking Water Wells," *Microb. Ecol.*, vol. 66, no. 1, pp. 19–29, 2013.

[33] M. L. Frisch, "Gaussian 09, Revision A.02." Gaussian, Inc, Wallingford CT, 2009.

[34] K. Krishnamoorthy, M. Veerapandian, K. Yun, and S. J. Kim, "The chemical and structural analysis of graphene oxide with different degrees of oxidation," *Carbon N. Y.*, vol. 53, pp. 38–49, 2013.

[35] Y. Zhang, S. Zhang, and T. S. Chung, "Nanometric Graphene Oxide Framework Membranes with Enhanced Heavy Metal Removal via Nanofiltration," *Environ. Sci. Technol.*, vol. 49, no. 16, pp. 10235–10242, 2015.

[36] V. N. Popov, L. Henrard, and P. Lambin, "Resonant Raman spectra of graphene with point defects," *Carbon N. Y.*, vol. 47, no. 10, pp. 2448–2455, 2009.

[37] A. C. Ferrari, "Raman spectroscopy of graphene and graphite: Disorder, electron-phonon coupling, doping and nonadiabatic effects," *Solid State Commun.*, vol. 143, no. 1–2, pp. 47–57, 2007.

[38] A. C. Ferrari and D. M. Basko, "Raman spectroscopy as a versatile tool for studying the properties of graphene," *Nat. Nanotechnol.*, vol. 8, no. 4, pp. 235–246, 2013.

[39] K. N. Kudin *et al.*, "Raman Spectra of Graphite Oxide and Functionalized Graphene Sheets," *Nano Lett.*, vol. 8, no. 1, pp. 36–41, 2007.

[40] S. Claramunt, A. Varea, D. López-Díaz, M. M. Velázquez, A. Cornet, and A. Cirera, "The importance of interbands on the interpretation of the raman spectrum of graphene oxide," *J. Phys. Chem. C*, vol. 119, no. 18, pp. 10123–10129, 2015.

[41] A. Sadezky, H. Muckenhuber, H. Grothe, R. Niessner, and U. Pöschl, "Raman microspectroscopy of soot and related carbonaceous materials: Spectral analysis and structural information," *Carbon N. Y.*, vol. 43, no. 8, pp. 1731–1742, 2005.

[42] R. Wu, Y. Wang, L. Chen, L. Huang, and Y. Chen, "Control of the oxidation level of graphene oxide for high efficiency polymer solar cells," *RSC Adv.*, vol. 5, no. 61, pp. 49182–49187, 2015.

[43] V. A. Sethuraman, L. J. Hardwick, V. Srinivasan, and R. Kostecki, "Surface structural disordering in graphite upon lithium intercalation/deintercalation," *J. Power Sources*, vol.

195, no. 11, pp. 3655–3660, 2010.

[44] K. G. Zhou *et al.*, "Electrically controlled water permeation through graphene oxide membranes," *Nature*, vol. 559, no. 7713, pp. 236–240, 2018.

[45] C. Vallés, F. Beckert, L. Burk, R. Mülhaupt, R. J. Young, and I. A. Kinloch, "Effect of the C/O ratio in graphene oxide materials on the reinforcement of epoxy-based nanocomposites," *J. Polym. Sci. Part B Polym. Phys.*, vol. 54, no. 2, pp. 281–291, 2016.

[46] M. Seredych, J. A. Rossin, and T. J. Bandosz, "Changes in graphite oxide texture and chemistry upon oxidation and reduction and their effect on adsorption of ammonia," *Carbon N. Y.*, vol. 49, no. 13, pp. 4392–4402, 2011.

[47] S. Stankovich, "Synthesis of graphene-based nanosheets via chemical reduction of exfoliated graphite oxide," *Carbon N. Y.*, vol. 45, pp. 1558–1565, 2007.

[48] H. Zhang *et al.*, "Carboxyl-functionalized graphene oxide polyamide nanofiltration membrane for desalination of dye solutions containing monovalent salt," *J. Memb. Sci.*, vol. 539, pp. 128–137, 2017.

[49] Z. Li *et al.*, "Uranium ( VI ) adsorption on graphene oxide nanosheets from aqueous solutions," *Chem. Eng. J.*, vol. 210, pp. 539–546, 2012.

[50] B. Konkena and S. Vasudevan, "Understanding aqueous dispersibility of graphene oxide and reduced graphene oxide through p K a measurements," *J. Phys. Chem. Lett.*, vol. 3, no. 7, pp. 867–872, 2012.

[51] ASTM, "Zeta potential of colloids in water and waste water. ASTM standar D 4187-82.," *Am. Soc. Test. Mater.*, 1985.

[52] H. Yan *et al.*, "Effects of the oxidation degree of graphene oxide on the adsorption of methylene blue," *J. Hazard. Mater.*, vol. 268, pp. 191–198, 2014.

[53] F. Baskoro *et al.*, "Graphene oxide-cation interaction: Inter-layer spacing and zeta potential changes in response to various salt solutions," *Journal of Membrane Science*, vol. 554. pp. 253–263, 2018.

[54] C.-J. Shih, S. Lin, R. Sharma, M. S. Strano, and D. Blankschtein, "Understanding the pH-Dependent Behavior of Graphene Oxide Aqueous Solutions: A Comparative Experimental and Molecular Dynamics Simulation Study," *Langmuir*, vol. 28, no. 1, pp. 235–241, 2011.

[55] M. Asadullah, I. Jahan, M. B. Ahmed, P. Adawiyah, N. H. Malek, and M. S. Rahman, "Preparation of microporous activated carbon and its modification for arsenic removal

from water," *J. Ind. Eng. Chem.*, vol. 20, no. 3, pp. 887–896, 2014.

[56] L. Lorenzen, J. S. J. van Deventer, and W. M. Landi, "Factors affecting the mechanism of the adsorption of arsenic species on activated carbon," *Miner. Eng.*, vol. 8, no. 4–5, pp. 557–569, 1995.

[57] Y. Bulut and H. Aydin, "A kinetics and thermodynamics study of methylene blue adsorption on wheat shells," *Desalination*, vol. 194, no. 1–3, pp. 259–267, 2006.

[58] B. H. Hameed, A. T. M. Din, and A. L. Ahmad, "Adsorption of methylene blue onto bamboo-based activated carbon: Kinetics and equilibrium studies," *J. Hazard. Mater.*, vol. 141, no. 3, pp. 819–825, 2007.

[59] K. B. Payne and T. M. Abdel-Fattah, "Adsorption of arsenate and arsenite by iron-treated activated carbon and zeolites: Effects of pH, temperature, and ionic strength," *J. Environ. Sci. Heal. - Part A Toxic/Hazardous Subst. Environ. Eng.*, vol. 40, no. 4, pp. 723–749, 2005.

[60] N. Kannan and M. M. Sundaram, "Kinetics and mechanism of removal of methylene blue by adsorption on various carbons - A comparative study," *Dye. Pigment.*, vol. 51, no. 1, pp. 25–40, 2001.

[61] M. M. Nassar and Y. H. Magdy, "Removal of different basic dyes from aqueous solutions by adsorption on palm-fruit bunch particles," *Chem. Eng. J.*, vol. 66, no. 3, pp. 223–226, 1997.

[62] A. O. Dada, A. P. Olalekan, A. M. Olatunya, and D. O., "Langmuir , Freundlich , Temkin and Dubinin – Radushkevich Isotherms Studies of Equilibrium Sorption of Zn 2 + Unto Phosphoric Acid Modified Rice Husk," *IOSR J. Appl. Chem.*, vol. 3, no. 1, pp. 38–45, 2012.

[63] C. Su and R. W. Puls, "Arsenate and arsenite removal by zerovalent iron: Effects of phosphate, silicate, carbonate, borate, sulfate, chromate, molybdate, and nitrate, relative to chloride," *Environ. Sci. Technol.*, vol. 35, no. 22, pp. 4562–4568, 2001.

[64] W. Wan, T. J. Pepping, T. Banerji, S. Chaudhari, and D. E. Giammar, "Effects of water chemistry on arsenic removal from drinking water by electrocoagulation," *Water Res.*, vol. 45, no. 1, pp. 384–392, 2011.

[65] X. Meng, S. Bang, and G. P. Korfiatis, "Effects of silicate, sulfate, and carbonate on arsenic removal by ferric chloride," *Water Res.*, vol. 34, no. 4, pp. 1255–1261, 2000.

[66] D. F. Rodrigues, "Investigation of acute effects of graphene oxide on wastewater microbial community : A case study Carbon-based nanomaterials – Flat sheets = graphene – Spherical and ellipsoidal = fullerenes," vol. 257, pp. 33–39, 2013.

[67] S. Liu *et al.*, "Antibacterial Activity of Graphite , Graphite Oxide , Graphene Oxide , and Reduced Graphene Oxide : Membrane and Oxidative Stress," *ASC Nano*, vol. 5, no. 9, pp. 6971–6980, 2011.